\documentclass[traditabstract]{aa} 
\usepackage{graphicx}
\usepackage{txfonts}

\def\ms{\hbox{\,m\,s$^{-1}$}}         
\def\m2s2{\hbox{\,m$^{2}$\,s$^{-2}$}} 
\def\kms{\hbox{\,km\,s$^{-1}$}}       
\def\vsini{\hbox{$v$\,sin\,$i$}}      
\def\sini{\hbox{sin\,$i$}}      

\begin{document}
\title{The HARPS search for southern extra-solar planets\thanks{Based on observations made with HARPS
spectrograph on the 3.6-m ESO telescope at La Silla Observatory under the GTO programme ID 072.C-0488. 
Tables of radial velocities are only available in electronic form at the CDS via anonymous ftp to 
cdsarc.u-strasbg.fr (130.79.128.5) or via http://cdsweb.u-strasbg.fr/cgi-bin/qcat?J/A+A/.}}

\subtitle{XVII. Super-Earth and Neptune-mass planets \\
in multiple planet systems HD\,47186 and HD\,181433}

\author{
Bouchy, F. \inst{1}
\and Mayor, M. \inst{2}
\and Lovis, C.\inst{2}
\and Udry, S. \inst{2}
\and Benz, W. \inst{3}
\and Bertaux, J.-L. \inst{4}
\and Delfosse, X. \inst{5}
\and Mordasini, C. \inst{3}
\and Pepe, F. \inst{2}
\and Queloz, D. \inst{2}
\and Segransan, D. \inst{2}
}

\offprints{\email{bouchy@iap.fr}}

\institute{
Institut d'Astrophysique de Paris, UMR7095 CNRS, Universit\'e Pierre \& Marie Curie, 98bis Bd Arago, 75014 Paris, France
\and
Observatoire de Gen\`eve, Universit\'e de Gen\`eve, 51 Ch. des Maillettes, 1290 Sauverny, Switzerland
\and
Physikalisches Institut Universitat Bern, Sidlerstrasse 5, 3012 Bern, Switzerland 
\and 
Service d'A\'eronomie du CNRS, BP 3, 91371 Verri\`eres-le-Buisson, France
\and 
Laboratoire d'Astrophysique, Observatoire de Grenoble, Universit\'e J. Fourier, BP 53, 38041 Grenoble, Cedex
9, France
}

\date{Received ; accepted }

\abstract
{This paper reports on the detection of two new multiple planet systems around solar-like stars 
HD\,47186 and HD\,181433. The first system includes a hot Neptune of 22.78 M$_{\oplus}$ at 4.08-days period and 
a Saturn of 0.35 M$_{\rm JUP}$ at 3.7-years period. The second system includes a Super-Earth of 7.5
M$_{\oplus}$ at 9.4-days period, a 0.64 M$_{\rm JUP}$ at 2.6-years period as well as a third companion
of 0.54 M$_{\rm JUP}$ with a period of about 6 years. These detections increase to 20 the number of close-in low-mass 
exoplanets (below 0.1 M$_{\rm JUP}$) and strengthen the fact that 
80\% of these planets are in a multiple planetary systems.}

\keywords{planetary systems -- Techniques: radial velocities -- stars: individual: HD\,47186 --
stars: individual: HD\,181433} 
               
\titlerunning{Hot-Neptunes in multiple planet systems HD\,47186 and HD\,181433}

\authorrunning{F. Bouchy et al.}

\maketitle
%

\section{Introduction}

The HARPS spectrograph based on the 3.6-m ESO telescope at La Silla Observatory is now in operation 
since 2003. The HARPS consortium started 5 years ago an ambitious and comprehensive Guaranteed Time 
Observations (GTO) program of high-precision systematic search for exoplanet in the Southern sky 
(Mayor et al. \cite{mayor03}, \cite{mayor08}). Relevant efforts were made for the search for very low-mass planets. 
Our consortium dedicates ~50\% of the GTO on HARPS since 2003 
to monitoring about 200 nearby non-active stars. Thank to radial velocity accuracy better \
than 1 {\ms} (Pepe et al. \cite{pepe04}, Lovis et al. \cite{lovis06b}) HARPS succeeds 
in the discovery of several Neptune-mass and Super-Earth exoplanets around solar-type stars 
(Santos et al. \cite{santos04}, Udry et al. \cite{udry06}, Lovis et al. \cite{lovis06}, Melo et al. 
\cite{melo08}, Mayor et al. \cite{mayor08}) and M dwarfs (Bonfils et al \cite{bonfils05}, 
Udry et al. \cite{udry07}, Bonfils et al. \cite{bonfils07}, Forveille et al. \cite{forveille08}). 
Such efficiency mainly comes from the following factors : 1) a careful and frequent monitoring 
of instrumental performances, 2) a continuous improvement 
of the data reduction software (e.g. Lovis \& Pepe \cite{lovis07}), 3) a dedicated and careful observing strategy 
to deal with stellar seismic noise (e.g. Bouchy et al. \cite{bouchy05}), 
4) a long-duration monitoring of non-active stars, 5) an accumulation of measurements in order to 
identify multi-planetary systems. Our large program is now leading to an increasing list of close-in 
low-mass exoplanets which will improve our knowledge of the planet distribution in the mass-period
diagram and will allow comparisons with theoretical predictions. Furthermore this increasing number of
close-in low-mass planets stimulates dedicated photometric follow-up in order to detect transiting 
Neptunes like GJ436b (Gillon et al. \cite{gillon07}). Indeed we expect statistically that 5-10\% of 
this close-in low-mass exoplanets offer the appropriate configuration to transit their parent stars. 
In that case a direct measurement of the planetary radius as well as the exact mass will be provided. 
In this paper, we present the discovery of two new multiple planet systems including one 
hot Neptune and one Super-Earth orbiting the stars HD\,47186 and HD\,181433 respectively.

\section{Parent star characteristics of HD\,47186 and HD\,181433.}

The basic photometric and astrometric properties of HD\,47186 and HD\,181433 
were taken from the Hipparcos catalogue (ESA \cite{esa97}). 
Accurate spectroscopic stellar parameters of the HARPS GTO ``high-precision'' program 
were determined by Sousa et al. (\cite{sousa08}) using the high-quality high-resolution and high S/N 
HARPS spectra. Stellar parameters for HD\,47186 and HD\,181433 are summarized in Table~\ref{table:1}. 

\begin{table}
\centering                       
\caption{Stellar parameters of  HD\,47186 and HD\,181433. The rotational period is derived from the 
activity index log\,$R^{'}_{HK}$.}             
\label{table:1}      
\begin{tabular}{l l l l}       
\hline\hline                 
Parameters & HD\,47186 & HD\,181433 & Reference  \\   
\hline      
Spectral type  & G5V & K3IV    & Hipparcos \\
Parallax [mas] & 26.43 & 38.24 & Hipparcos \\
Distance [pc]  & 37.84 & 26.15 & Hipparcos \\
$m_v$          & 7.6   &  8.4  & Hipparcos \\
B-V            & 0.71  &  1.01 & Hipparcos \\
$M_v$          & 4.74  &  6.31 & Hipparcos \\
Luminosity [L$_\odot$]  & 1.08$\pm$0.029 & 0.308$\pm$0.026 & Sousa et al. (\cite{sousa08}) \\
Mass [M$_\odot$]  & 0.99 & 0.78 &  Sousa et al. (2008) \\
$T_{\rm eff}$ [K]  & 5675$\pm$21   & 4962$\pm$134 & Sousa et al. (\cite{sousa08}) \\
log\,$g$         & 4.36$\pm$0.04 &  4.37$\pm$0.26 & Sousa et al. (\cite{sousa08}) \\
$[Fe/H]$         & 0.23$\pm$0.02 & 0.33$\pm$0.13  & Sousa et al. (\cite{sousa08}) \\
{\vsini} [\kms]  & 2.2 & 1.5 & this paper \\
log\,$R^{'}_{HK}$  & -5.01 & -5.11 & this paper \\
P$_{\rm ROT}$ [days]  & 33  & 54 & this paper \\
\hline                                   
\end{tabular}
\end{table}

\section{Radial-velocity data and orbital solutions}

The observations were carried out using the HARPS spectrograph (3.6-m ESO telescope, La Silla, Chile). 
We have obtained 66 and 107 measurements on HD\,47186 and HD\,181433 respectively spanning 
more than 4 years. The exposure time was fixed to 900s in order to average out the stellar seismic noise. 
The spectra have typical S/N per pixel in the range 120-250 for HD\,47186 and 80-160 for HD\,181433 
reflecting the difference in magnitude. Radial velocities (available in electronic form at CDS) were 
obtained directly using the HARPS pipeline. Their uncertainties, including photon noise, wavelength 
calibration uncertainty and spectrograph drift uncertainty, are in the range 0.3-0.6 {\ms} 
and 0.4-1.0 {\ms} for HD\,47186 and HD\,181433 respectively. 

\subsection{HD\,47186}

\begin{figure}
\centering
\includegraphics[width=8.5cm]{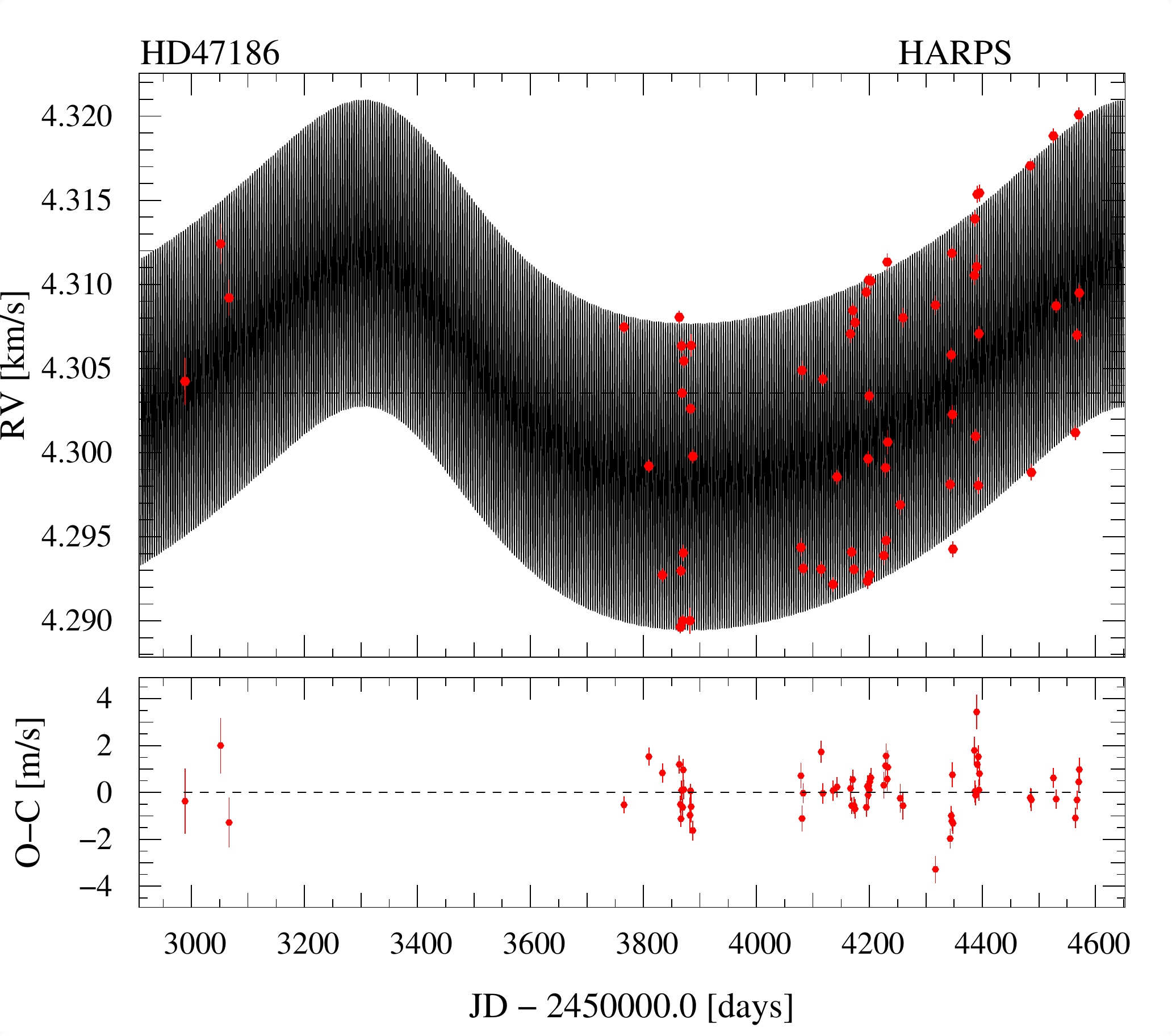}
\includegraphics[width=8.5cm]{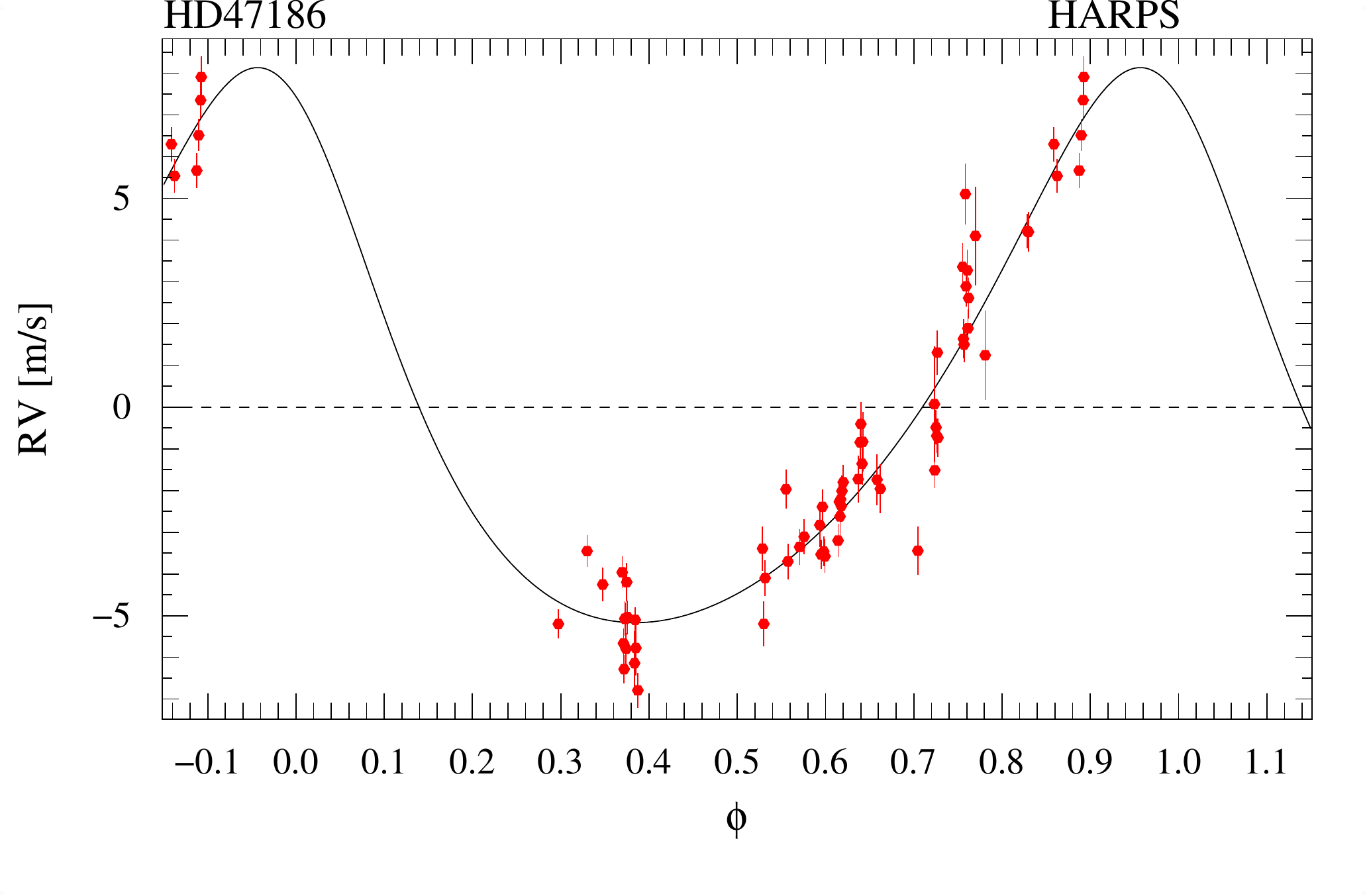}
\includegraphics[width=8.5cm]{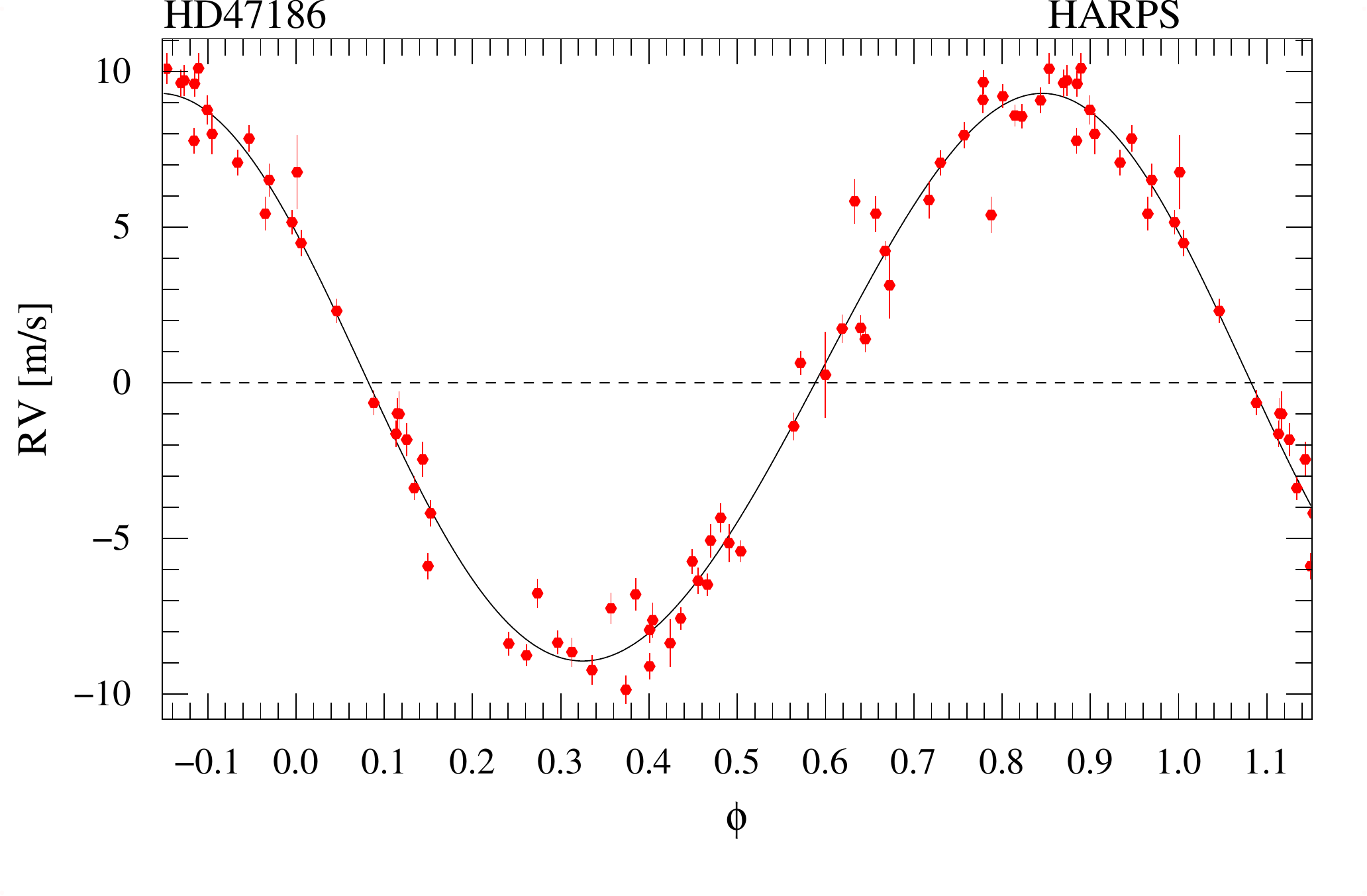}
\caption{2-planet Keplerian model for the HD\,47186 radial velocity variations. 
The upper panel display the RVs as a function of Julian Date. The middle and bottom 
panels display the phase-folded curve of the 3.7-years and the 4-days period planet 
respectively, with points representing the observed RVs, after removing the effect of 
the other planets.}
\label{rv_hd47186}
\end{figure}

Radial velocity measurements of HD\,47186 as a function of Julian Date are shown on top of 
Fig.~\ref{rv_hd47186}. Analysis of these data reveals the presence of a clear and 
stable 4-day period signal in addition with a long-term modulation of 3.7-year period. 
The phase-folded curves of the two planets, with points representing the observed radial 
velocities, after removing the effect of the other planets are displayed in 
Fig.~\ref{rv_hd47186} (middle and bottom panels). The reduced $\chi^2$ per degree of freedom 
is 2.25 and the residuals around the solution is 0.91 \ms. The derived orbital parameters 
lead to a minimum mass of 22.8 M$_{\oplus}$ and a separation $a=0.05$ AU for the close-in 
exoplanet and a minimum mass of 0.35 M$_{\rm JUP}$ and a separation $a=2.4$ AU for the 
second exoplanet. Orbital and physical parameters derived from the 2-planet Keplerian 
models are presented in Table~\ref{orbit1}. The close-in planet has a slightly significant 
eccentricity (0.038$\pm$0.020). If we fix to zero the eccentricity, all parameters keep 
the same except the periastron epoch T$_{\rm peri}$ = 54566.261$\pm$0.028. In that case, 
the reduced $\chi^2$ is 2.28 and the residuals around the solution is 0.94 \ms. 
Orbital parameters of the long-period 
planets may still be improved. Although the fact that more than 1 orbital period was covered, 
no measurements were made in between phases from -0.1 to 0.3 and we assumed here that there 
is no long-term trend. Additional measurements will significantly improved 
the determination of the orbital parameters and will put constraints on 
a possible third companion at longer period.  
We checked that the bisector shape of the cross-correlation function (see Queloz et al. 2001) 
shows no variations down to the photon noise level, giving strong support to the 
planetary interpretation of the 2 RV signals. 


\begin{table}
\centering                       
\caption{Orbital and physical parameters of the 2-planet system orbiting HD\,47186.}             
\label{orbit1}      
\begin{tabular}{l c c}       
\hline\hline                 
Parameters & HD\,47186b & HD\,47186c \\   
\hline      
P [days]          & 4.0845$\pm$0.0002 & 1353.6$\pm$57.1 \\
T$_{\rm peri}$ [BJD-2400000]   & 54566.95$\pm$0.36 & 52010$\pm$180  \\
e                 & 0.038$\pm$0.020   & 0.249$\pm$ 0.073 \\
$\omega$ [deg]    & 59$\pm$32         &  26$\pm$23     \\
V [\kms]          & \multicolumn{2}{c}{4.3035$\pm$0.0014}   \\
$K$ [\ms]         & 9.12$\pm$0.18     &  6.65$\pm$1.43 \\
m{\sini} [$M_{\rm JUP}$]   &  0.07167 & 0.35061        \\
m{\sini} [$M_{\oplus}$]    &  22.78   & 111.42      \\
$a$ [AU]                  & 0.050    &   2.395    \\
\hline   
$N_{\rm meas}$           & \multicolumn{2}{c}{66}   \\
Data span [days]            & \multicolumn{2}{c}{1583}  \\
$\sigma$ (O-C) [\ms]     & \multicolumn{2}{c}{0.91}  \\
$\chi^2_{\rm red}$       & \multicolumn{2}{c}{2.25}  \\ 			     
\hline
\end{tabular}
\end{table}

\subsection{HD\,181433}

\begin{figure}
\centering
\includegraphics[width=8.5cm]{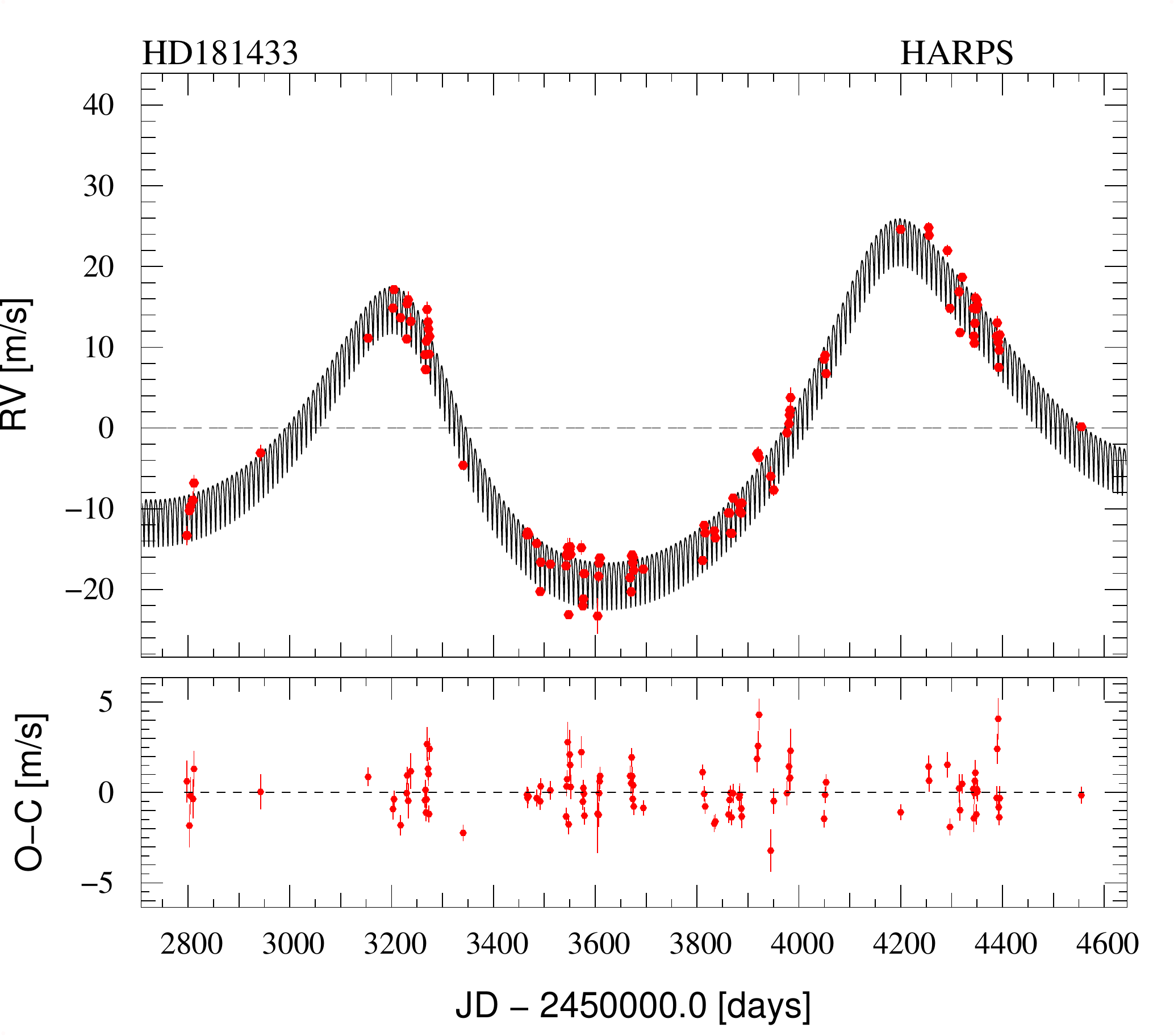}
\includegraphics[width=8.5cm]{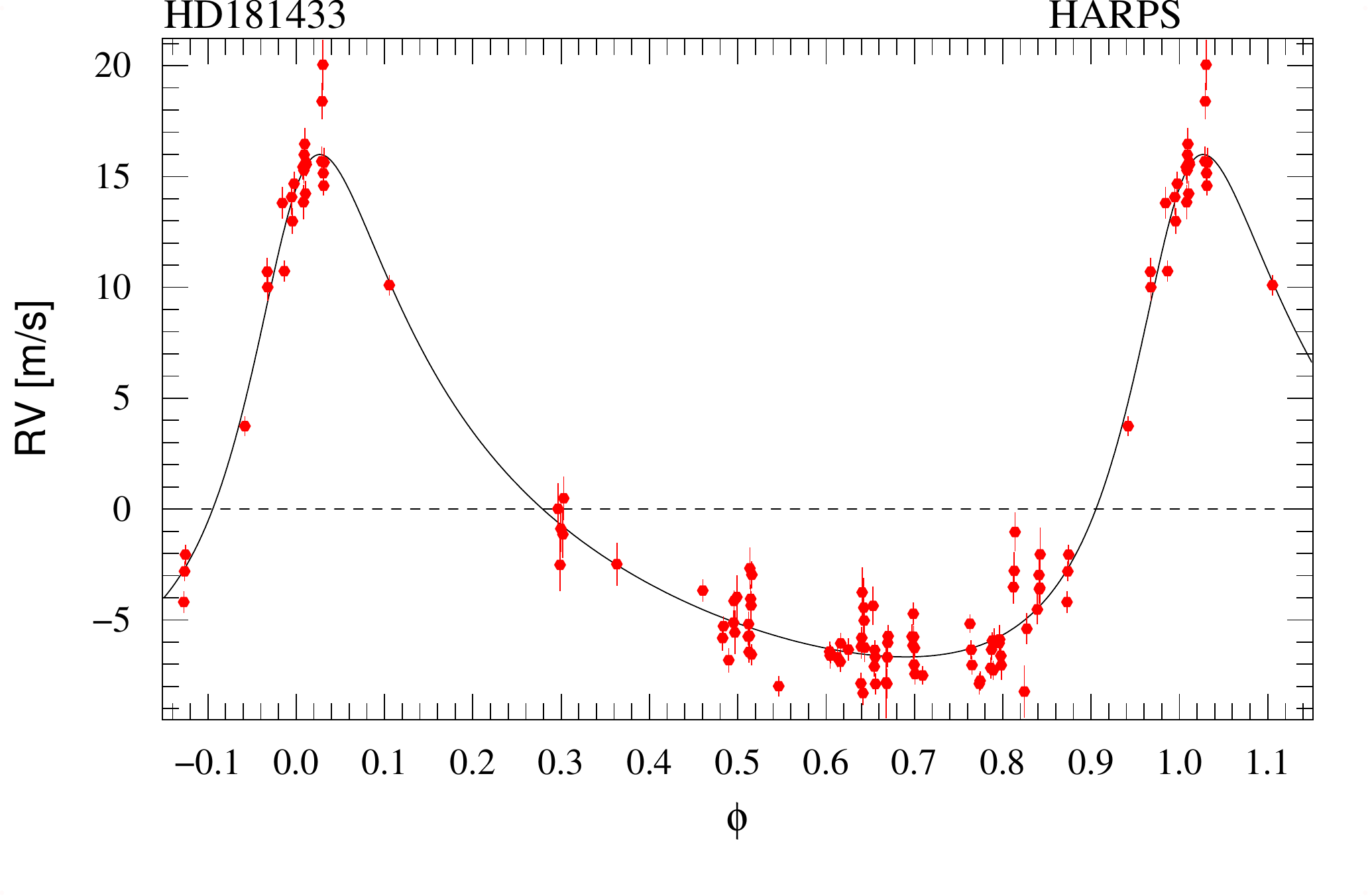}
\includegraphics[width=8.5cm]{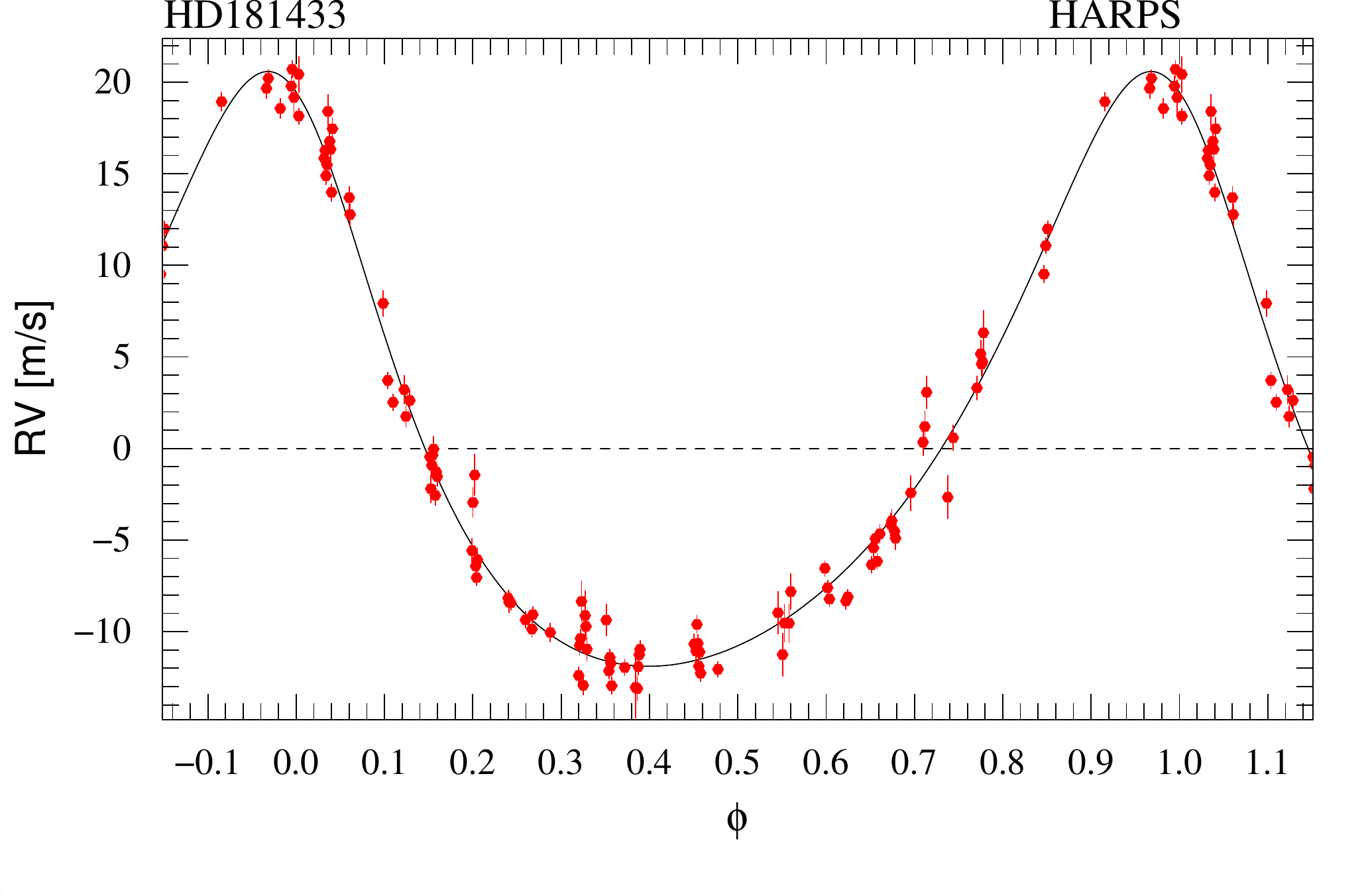}
\includegraphics[width=8.5cm]{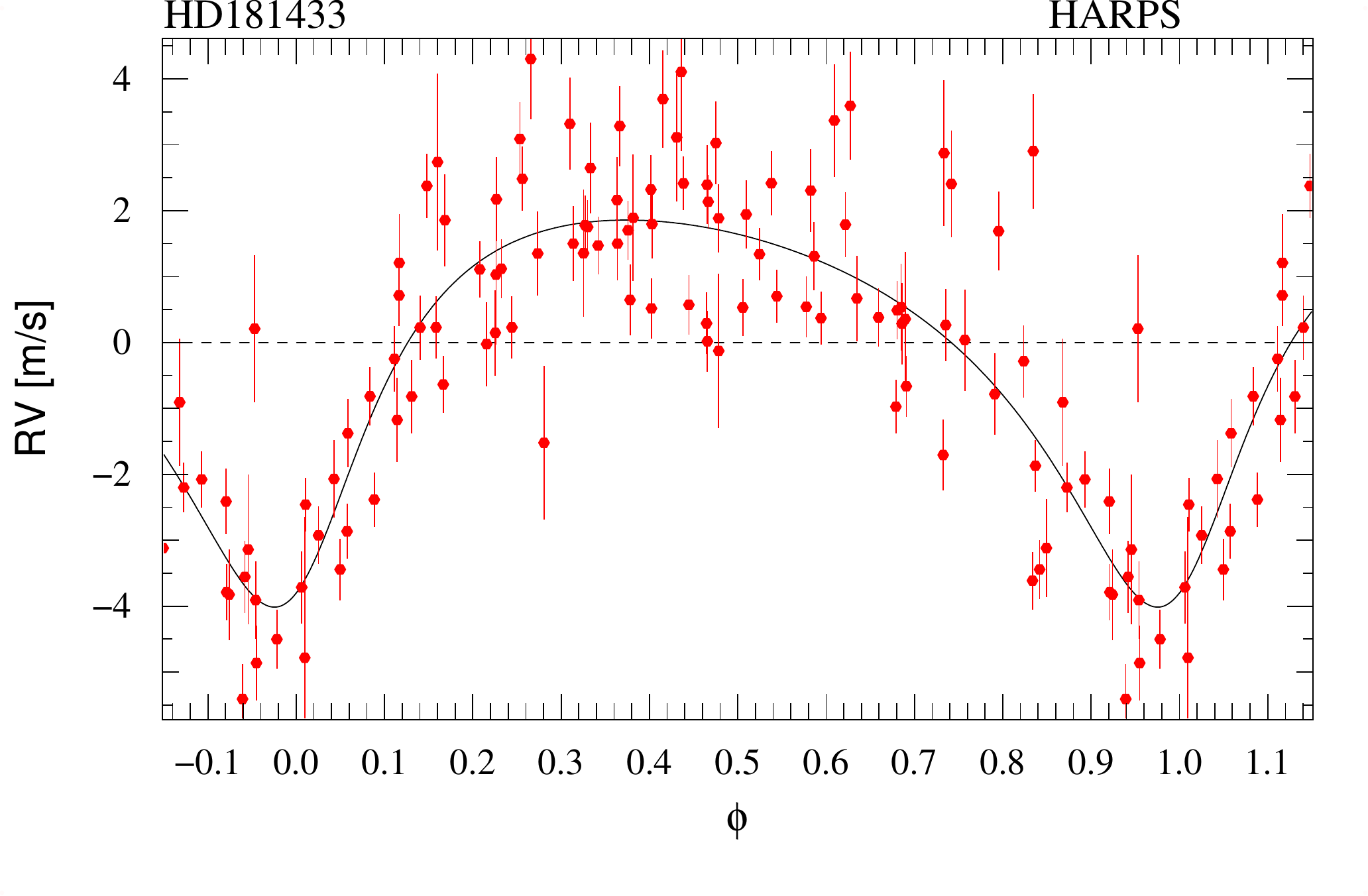}
\caption{3-planet Keplerian model for the HD\,181433 radial velocity variations. 
The upper panel display the RVs as a function of Julian Date. The 3 bottom 
panels (from top to bottom) display the phase-folded curve of the 6-years, 2.6-years and the 9.4-days 
period planets respectively, with points representing the observed RVs, after removing the effect of 
the other planets.}
\label{rv_hd181433}
\end{figure}

\begin{figure}
\centering
\includegraphics[width=8.5cm]{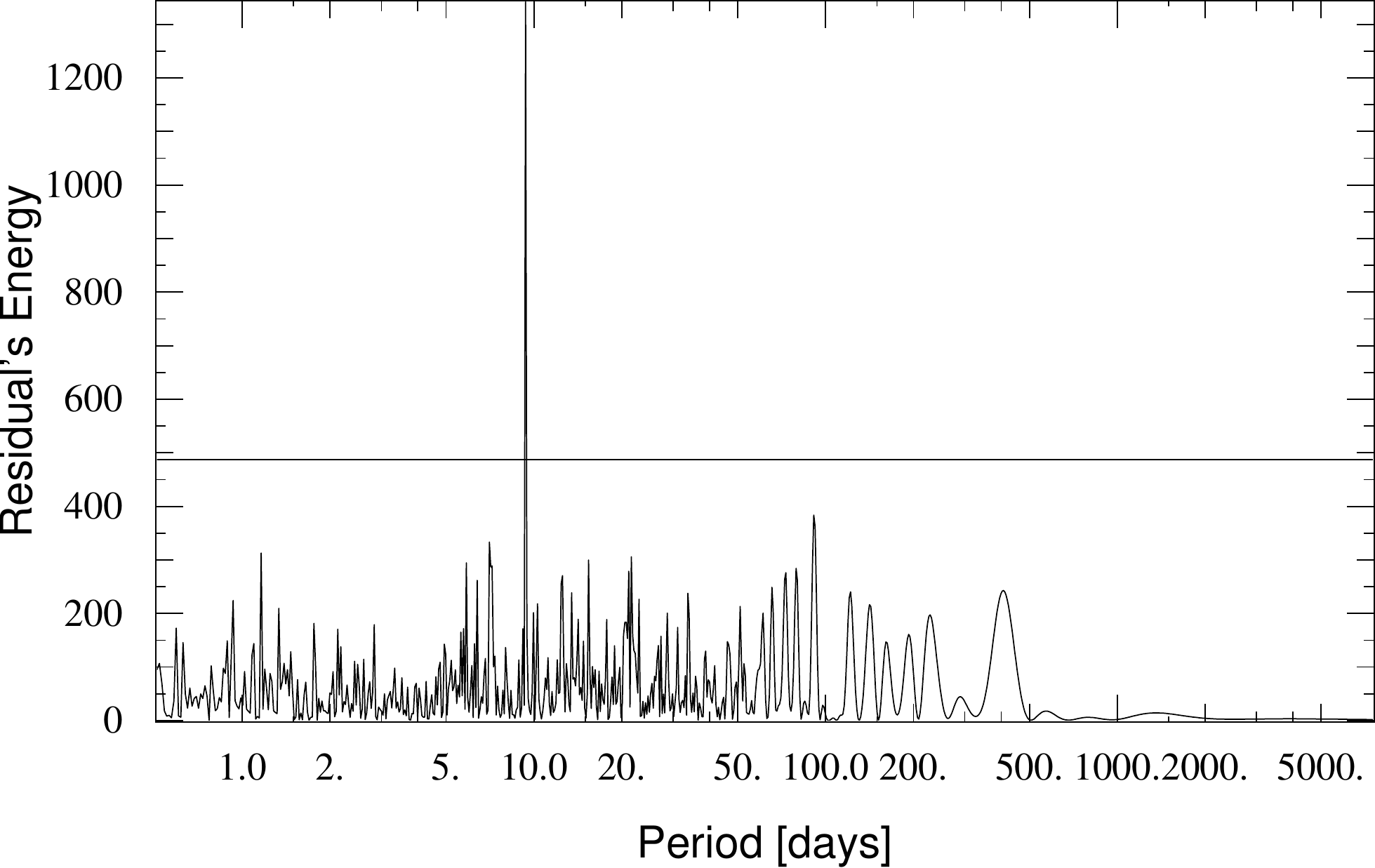}
\caption{Lomb-Scargle periodogram of the radial velocities of HD181433 after substraction 
of the two long period signals. The horizontal line corresponds to a false 
alarm probability of 10$^-4$.} 
\label{fap}
\end{figure}

Our radial velocity measurements as a function of Julian Date are shown on top of
Fig.~\ref{rv_hd181433}. It first shows a clear signal at period 2.6 years plus an additional 
long-term trend. Analysis of the residuals reveals an additional signal at 9.4 days.       
We tested a linear, a quadratic and a Keplerian solution for the long-term trend. 
It does not affect the solution for the short-period planet but affect slightly the orbital 
parameters of the 2.6 years planet. With a linear trend of 1.70$\pm$0.26 \ms/year, 
the reduced $\chi^2$  is 3.3 and the residuals around the solution is 1.66 \ms. A quadratic 
trend with a slope of 2.98$\pm$0.38 \ms/year and a curve of 1.30$\pm$0.22 \ms/year$^2$ improves 
the global solution and gives a reduced $\chi^2$ of 2.4 and residuals of 1.22 \ms. 
Finally a Keplerian orbit with a period of about 6 years improves significantly the global 
solution and gives a reduced $\chi^2$ of 1.3 and residuals of 1.06 \ms. The parameters 
of this long period planet is not well constrained by our span coverage of 4.8 years. 
However long term observations of HD181433 carried with the spectrograph CORALIE 
over 9 years confirm this signal.
The phase-folded curves of the three planets, with points representing the observed radial 
velocities, after removing the effect of the other planets are displayed in Fig.~\ref{rv_hd181433}. 
The derived orbital parameters 
lead to a minimum mass of 7.5 M$_{\oplus}$ and a separation $a=0.08$ AU for the close-in 
exoplanets and a minimum mass of 0.64 M$_{\rm JUP}$ and a separation $a=1.76$ AU for the 
second exoplanets. The derived parameters of the third exoplanet indicate a minimum 
mass of 0.54 M$_{\rm JUP}$ and a semi-major axis close to 3 AU. Additional CORALIE 
observations and dynamical analysis of this system will be presented in a forthcoming paper.    
Orbital and physical parameters derived from the 3-planet Keplerian 
models are presented in Table~\ref{orbit2}. We checked that the bisector shape of the CCF shows 
no variations and no correlation with any of the 3 RV signals. The significant of the 9.4-days  
planet was checked using a Monte Carlo approach in which the residuals to the two external  
planets fit are scrambled and then analyzed in order to find periodicity. 
Figure~\ref{fap} represents the Lomb-Scargle periodiogram of these residuals. 
The 9.4-day signal appears strongly upper the false alarm probability limit.

\begin{table}
\centering                       
\caption{Orbital and physical parameters of the 3-planet system orbiting HD\,181433.}             
\label{orbit2}      
\begin{tabular}{l c c c}       
\hline\hline                 
Parameters & HD\,181433b & HD\,181433c & HD\,181433d \\   
\hline      
P [days]           		& 9.3743$\pm$0.0019 	& 962.0$\pm$15 		& 2172$\pm$158    \\
T$_{\rm peri}$ [BJD-2400000]    & 54542.0$\pm$0.26 	& 53235.0$\pm$7.3	& 52154$\pm$194 \\
e                 		& 0.396$\pm$0.062	& 0.28$\pm$0.02		& 0.48$\pm$0.05 \\
$\omega$ [deg]       		& 202$\pm$10 		& 21.4$\pm$3.2 		& -30$\pm$13 \\
V [\kms]          & \multicolumn{3}{c}{40.2125$\pm$0.0004} \\
$K$ [\ms]          		&  2.94$\pm$0.23 	& 16.2$\pm$0.4		& 11.3$\pm$0.9 \\
m{\sini} [$M_{\rm JUP}$]      	& 0.024 		& 0.64 			& 0.54 \\
m{\sini} [$M_{\oplus}$]         & 7.5 			& 203 			& 171  \\
$a$ [AU]                        & 0.080 		& 1.76 			& 3 \\
\hline   
$N_{\rm meas}$          & \multicolumn{3}{c}{107} \\
Data span [days]           & \multicolumn{3}{c}{1757} \\
$\sigma$ (O-C) [\ms]    & \multicolumn{3}{c}{1.06} \\
$\chi^2_{\rm red}$      & \multicolumn{3}{c}{1.29} \\				   
\hline
\end{tabular}
\end{table}

\section{Discussion and conclusion}

\begin{figure}
\centering
\includegraphics[width=8.5cm]{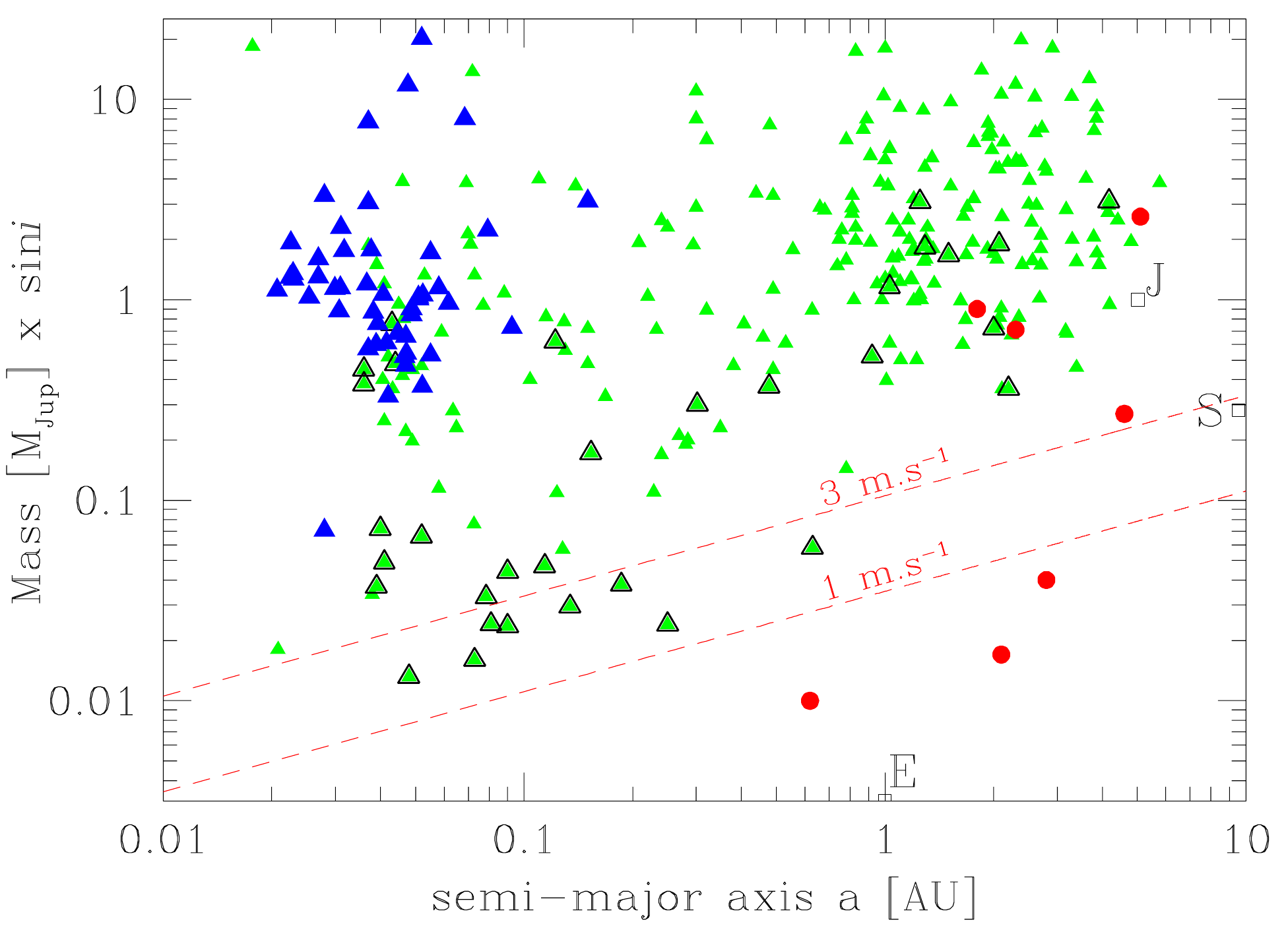}
\caption{Mass-separation diagram of the 300 known exoplanets. The triangles 
refer to exoplanets found by radial velocities. The dark triangles refer to transiting exoplanets. 
The circles refer to exoplanets found by microlensing. The bold triangles correspond to HARPS 
discovered exoplanets. Lines of radial-velocity semi-amplitude of 1 and 3 {\ms} are added 
assuming a 1 solar-mass star.}
\label{ma_diag}
\end{figure}

Figure~\ref{ma_diag} represents the $\sim$300 known exoplanets\footnote{from The Extrasolar Planets 
Encyclopedia (http://exoplanet.eu) June 2008.} in the mass-separation diagram. The triangles 
refer to exoplanets found by radial velocities. The dark triangles refer to transiting exoplanets. 
The circles refer to exoplanets found by microlensing. The bold triangles correspond to HARPS 
discovered exoplanets including the 5 ones described in this paper and the 3 ones orbiting HD\,40307 
(Mayor et al. 2008). HARPS is presently filling the area of close-in low-mass exoplanets with minimum 
mass below 0.1 M$_{\rm JUP}$. One expects that few \% of these exoplanets offer the 
appropriate configuration to transit their parent stars. In that case a direct measurement of the 
planetary radius as well as the exact mass will be done providing crutial informations and 
constraints about their composition. This was the case of the up-to-day 
unique transiting Neptune GJ436b (Gillon et al. \cite{gillon07}) which was first discovered 
by radial velocity (Butler et al. \cite{butler04}). Such close-in low-mass planets are expected to have 
radius of few R$_\oplus$ which means transit depth in the sub-millimagnitude regime. 
Although impossible to be detected in ground-based photometry surveys, such small transiting 
exoplanets may be detected in specific high-precision ground-based photometric follow-up, 
especially in case of small stellar radius (K and M dwarves), or using spaced-based facilities.
One advantage comes from the fact that these planets orbit around nearby bright stars 
which makes possible deep and further characterization. 

The two new planetary systems described here strengthen the fact low-mass exoplanets are found 
in multiple planetary systems. Indeed, from the 20 known exoplanets with masses lower than 
0.1 M$_{\rm JUP}$, 16 are in a multiple planetary systems, hence 80\%. This fraction should be 
compared to the 23\% of the $\sim$300 known exoplanets which are in a multiple planetary systems. 
If we restrict the analysis to planetary period shorter than 50 days, which correspond 
to the detectability period cutoff of low mass explanets, the conclusion remains the same, 
19\% of exoplanets and 72\% of low mass explanets are in a multiple systems. 
We note that from the 4 low-mass planets which are not up-to-day identified as part of a multiple 
planetary systems, 3 are orbiting a M-dwarf (GJ674b, GJ436b and HD\,285968b). 

\begin{figure}
\centering
\includegraphics[width=8cm]{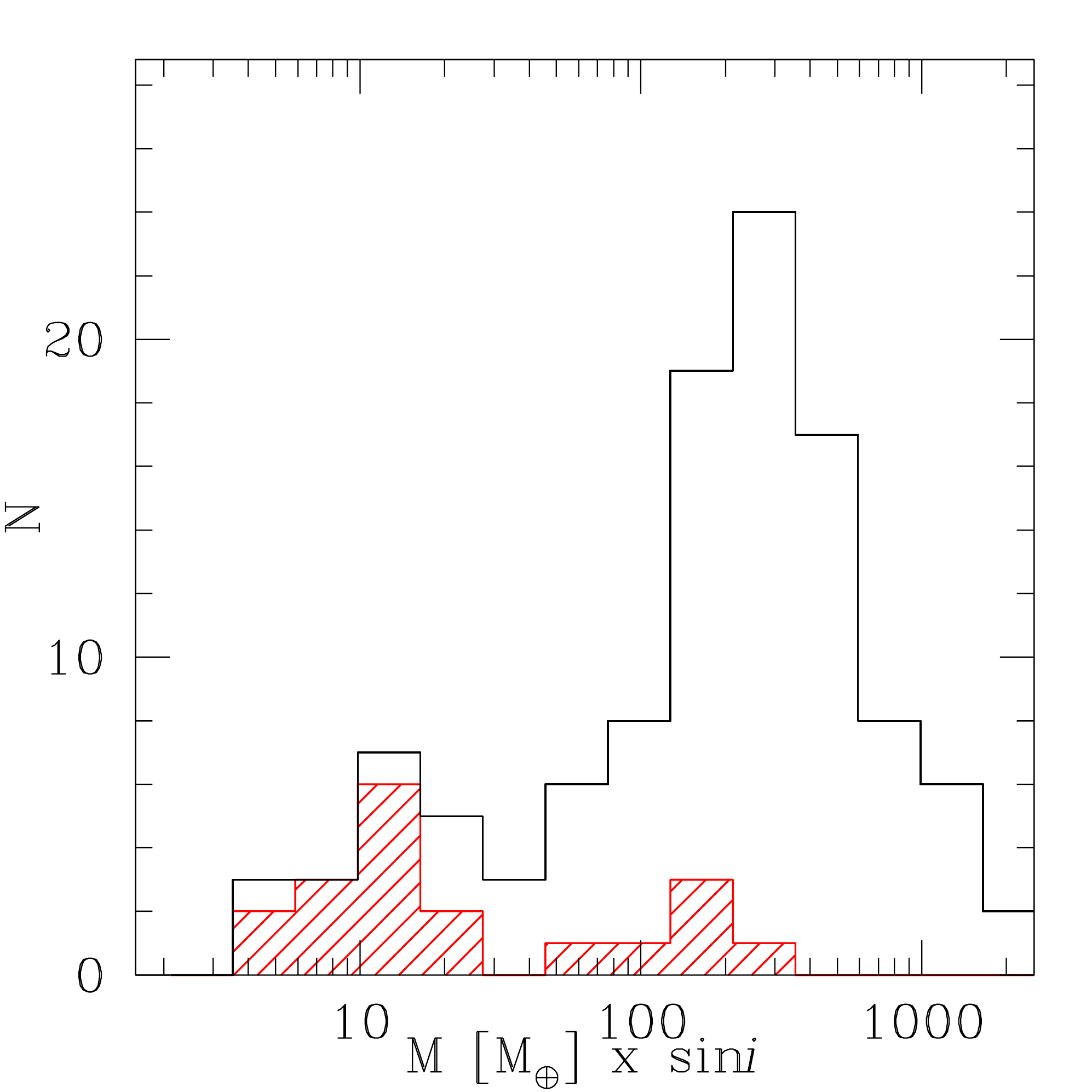}
\caption{Distribution of planetary masses from giant gaseous planets to the super-Earth 
regime for close-in planets (P$\le$50 days). The hatched histogram represents HARPS detections.}
\label{bimodal}
\end{figure}

It is interesting to discuss the discoveries presented in this paper  
from a theoretical point of view. In the planet population synthesis  
calculations of Mordasini et al. (\cite{mordasini08a}) based on the core accretion  
scenario, a large number of hot-Neptunian and Super-Earth planets as  
the ones presented here are predicted to exist at small distances from  
the host star. These low mass planets form a distinct  
sub-population of planets which did not undergo gas runaway accretion,  
in contrast to the sub-population of Hot Jupiters. In the predicted IMF 
of the close-in planets, a local minimum is found between the two groups, 
occurring  for solar type stars at a mass of about 30 M$_{\oplus}$.
Indeed, Fig.~\ref{ma_diag} indicates that the high precision program of HARPS has  
started to explore this new sub-population of close-in low mass  
planets. One might even tentatively see in Fig.~\ref{ma_diag} a bimodal mass 
distribution of the ``Hot'' planets. 
This emerging bimodal shape of the mass distribution from gaseous giant planets 
to the super-Earth regime was already pointed out by Mayor \& Udry (\cite{mayor08b}) 
and is also visible on the mass distribution of detected planet with period shorter than 
50 days (see Fig.~\ref{bimodal}). Planetary formation simulations also based on the core 
accretion scenario and done by Ida \& Lin (\cite{ida04a}, \cite{ida08}) 
also predict a bimodal distribution from gaseous giants to Super-Earth and furthermore 
a paucity of extrasolar planets with mass in the range 10-100  M$_{\oplus}$.

Discoveries of Hot Neptunes and close-in Super-Earth planets raise  
questions about how they were exactly formed. 
In-situ formation seems very unlikely for HD\,47186b. This is due to the  
fact that in-situ formation allows the assembly of planets of only a  
few times the Earth mass even at the supersolar metallicity of about  
0.2-0.3 dex of the stars considered here. HD\,47186b has in  
contrast a clearly larger, Neptunian like mass and is located at a  
very small semi-major axis, where only tiny amounts of planetary  
building blocks are available, if any (due to evaporation of solids  
very close to the star). For HD\,181433b, the planetary mass is only about a third, while the  
semi-major axis is larger, so that its mass is no more so clearly out  
of reach of what might can form in-situ. Still, for reasonable disk  
masses and profiles, it is unlikely that a such a massive Super-Earth  
planet can form at its current location.
Therefore, it seem necessary that some migration process was at work.  
Various processes can bring planets close to the star: Planet-disc  
interaction in the form of type I and II migration (e.g. Terquem \& Papaloizou \cite{terquem07}), 
planet-planet interaction in the form of shepherding (e.g. Mandell et al. \cite{mandell07}), 
scattering with subsequent circularization or the Kozai mechanism, among some others (e.g. Raymond  
et al. \cite{raymond08}; Zhou \& Lin \cite{zhou08}).

It is interesting to discuss the discoveries in the context of the two  
competing giant planet formation mechanisms. In the core accretion  
mechanism, the formation of a system with both giant and low mass  
(Neptunian or Super-Earth) planets can be regarded as a natural  
outcome, which is not necessary the case in the gravitational  
instability model. Especially a planetary system architecture with low  
mass planets at smaller semi-major axes and one or several giant  
planets at larger distances is expected in the baseline core accretion  
model without long distance migration, as the increase of the  
available planetary building blocks with distance facilitates giant  
planet growth at larger distances, while the smaller amounts available  
at shorter distances should lead to the formation of low mass planets.
This simple picture is reflected in both HD\,47186 and HD\,181433, even if  
the masses of the inner planets are large compared to those of the  
Solar System. We therefore may regard these systems as cases where  
some, but not extreme migration of both the low mass planets and the  
giants occurred. This would indicate an initial disk mass larger than  
that of the Solar System, but less massive than one leading to the  
formation of Hot Jupiters.

Population synthesis calculations based on the core accretion paradigm  
reproduce the ''metallicity effect'' (Ida \& Lin \cite{ida04b}) i.e. the strong 
positive correlation between the stellar metallicity and the planetary detection 
probability. Recent discoveries mainly by HARPS   
have shown that this ``metallicity effect'' might not exists for  
close-in low mass planets, even though the data set is currently too  
small for definitive conclusions (Udry \&  Santos \cite{udry07b}).
However the multiple planetary systems like HD\,47186 and HD\,181433 with both 
a giant and a low mass planet (55 Cnc, Gl876, HD\,160691, HD\,190360, HD\,219828) 
have all supersolar metallicities. In the core  
accretion scenario this is understood in the following way: High  
[Fe/H] stars (disks) are able to produce both giant and low mass  
planets, while low [Fe/H] systems are able to produce low mass planets  
only.  More discoveries of low mass planets will help clarify such  
correlations.

\begin{acknowledgements}
The authors thank the different observers from the other HARPS GTO sub-programmes 
who have also measured HD\,47186 and HD\,181433. We are grateful to all the staff of 
La Silla Observatory for their contribution to the success of the HARPS project. We wish 
to thank the Programme National de Plan\'etologie (INSU-PNP) and the Swiss National Science 
Foundation for their continuous support to our planet-search programs.    
\end{acknowledgements}

\end{document}